\documentclass[aps,prb,reprint,noeprint,superscriptaddress]{revtex4-2}
\pdfoutput=1
\usepackage[utf8]{inputenc}
\usepackage[english]{babel}
\usepackage{microtype}
\usepackage{amsmath,amssymb,amsfonts}
\usepackage{braket}
\usepackage{bm}
\usepackage{graphicx}
\usepackage{booktabs}

\usepackage[
    pdftitle={},
    pdfauthor={},
    colorlinks=true,
    unicode=true,
    pdfborder={0 0 0},
    allcolors=blue
]{hyperref}

\renewcommand{\vec}[1]{\bm{#1}}
\newcommand{\im}{\mathrm{i}}
\newcommand{\e}{\mathrm{e}}
\renewcommand{\Im}{\operatorname{\mathsf{Im}}}

\newcommand{\vk}{{\vec k}}
\newcommand{\VV}[2]{\begin{pmatrix}#1\\#2\end{pmatrix}}

\newcommand{\MM}[4]{\begin{pmatrix}#1 & #2 \\ #3 & #4\end{pmatrix}}

\newcommand{\cc}{c_{-\vk\downarrow}c_{\vk\uparrow}}
\newcommand{\cDcD}{c_{\vk\uparrow}^\dagger c_{-\vk\downarrow}^\dagger}
\newcommand{\ccD}{c_{-\vk\downarrow} c_{-\vk\downarrow}^\dagger}
\newcommand{\cDc}{c_{\vk\uparrow}^\dagger c_{\vk\uparrow}}
\newcommand{\aDa}{\alpha_{\vk}^\dagger\alpha_{\vk}}
\newcommand{\bDb}{\beta_{\vk}^\dagger\beta_{\vk}}
\newcommand{\ab}{\alpha_{\vk}\beta_{\vk}}
\newcommand{\aDbD}{\alpha_{\vk}^\dagger\beta_{\vk}^\dagger}

\begin{document}

\title{Momentum-resolved analysis of condensate dynamic and Higgs oscillations\\
in quenched superconductors with tr-ARPES}

\author{Lukas Schwarz}
\affiliation{Max Planck Institute for Solid State Research,
70569 Stuttgart, Germany}

\author{Benedikt Fauseweh}
\affiliation{Theoretical Division, Los Alamos National Laboratory,
Los Alamos, New Mexico 87545, USA}

\author{Dirk Manske}
\affiliation{Max Planck Institute for Solid State Research,
70569 Stuttgart, Germany}

\date{\today}

\begin{abstract}
Higgs oscillations in nonequilibrium superconductors provide an unique tool
to obtain information about the underlying order parameter.
Several properties like the absolute value, existence of multiple gaps
and the symmetry of the order parameter
can be encoded in the Higgs oscillation spectrum.
Studying Higgs oscillations with time-resolved
angle-resolved photoemission spectroscopy (ARPES)
has the advantage over optical measurements
that a momentum-resolved analysis of the condensate dynamic is possible.
In this paper,
we investigate the time-resolved spectral function measured in ARPES
for different quench protocols.
We find that analyzing amplitude oscillations of the ARPES intensity
in the whole Brillouin zone allows to understand how the condensate dynamic
contributes to the emerging of collective Higgs oscillations.
Furthermore,
by evaluating the phase of these oscillations
the symmetry deformation dynamic of the condensate can be revealed,
which gives insight about the ground state symmetry of the system.
With such an analysis,
time-resolved ARPES experiments might be used in future
as a powerful tool in the field of Higgs spectroscopy.
\end{abstract}

\maketitle

\section{Introduction}
Angle-resolved photoemission spectroscopy (ARPES) is a powerful method
as it allows to measure the electronic structure of materials
directly \cite{RevModPhys.75.473}.
Combined with additional time resolution (tr-ARPES),
dynamic processes of systems out of equilibrium can be studied in great detail
\cite{PhysRevLett.102.136401}.
Hereby, a pump pulse excites the system in a nonequilibrium state
prior to the photoemission probe pulse,
which is applied after a variable time delay
to scan the dynamic of the system.

In recent years, there was an evolving interest
in studying collective excitations of materials in nonequilibrium
as these can provide fundamental insights
into the internal symmetries and properties of a system.
This has been demonstrated on charge-density wave systems
\cite{PhysRevLett.97.067402,PhysRevLett.107.177402,Nature.471.490}
and high-temperature cuprate superconductors
\cite{PhysRevLett.99.197001,NatPhys.7.805,NatMater.17.416}.
In superconductors, an intrinsic collective amplitude (Higgs) mode
of the order parameter exists due to spontaneous $U(1)$ symmetry breaking
at the transition to the superconducting state
\cite{JLowTempPhys.126.901,annurev.varma2015}.
Exciting the Higgs mode is challenging
and only in the last years an exploration
of this area has started.
This progress was heavily supported by the upcoming of new technologies
like ultrafast THz spectroscopy
\cite{PhysicsToday.65.44},
which possess the required excitation energy
to not fully deplete the superconducting condensate in the pump process.

The main reason for the difficulties to excite the Higgs mode
is due to the fact
that the Higgs mode is a scalar mode
with neither net charge nor electric dipole,
which requires to go beyond the linear excitation regime.
While the first predictions of the Higgs mode
in superconductors are relatively old
\cite{PhysRevLett.47.811,PhysRevB.26.4883},
only indirect measurements in systems with competing charge-density wave orders
could be realized in the beginning
\cite{PhysRevLett.45.660}.
The first direct observation was performed in a THz pump-probe experiment
on the $s$-wave superconductor Nb$_{1-x}$Ti$_{x}$N
\cite{PhysRevLett.111.057002},
where Higgs oscillations of the order parameter,
reflected in oscillations of the electromagnetic response, could be observed.
Since then, only a few more experiments were reported
\cite{PNAS.110.4539,Science.345.1145,%
PhysRevB.96.020505,PhysRevLett.120.117001,NatCommun.11.1793},
but theoretical works on the subject became more popular
as it became clear that Higgs oscillations in nonequilibrium systems
provide rich information about the superconducting ground state.

It was found that much information about the ground state order parameter
is encoded in the Higgs oscillation frequency.
First of all, the main Higgs oscillation frequency
corresponds to the maximum of the absolute value of the order parameter itself
\cite{SovPhysJETP.38.1018,PhysRevLett.96.097005,PhysRevLett.96.230404,%
PhysRevLett.115.257001,PhysRevB.95.104507}.
In addition, multiband systems show frequencies for each band
and also for the frequency of the relative phase mode (Leggett mode)
\cite{NatCommun.7.11921,PhysRevB.95.104503}.
Excited in an asymmetric way,
nontrivial gap symmetries can show additional oscillation frequencies
as a result of an oscillation of the condensate
in a different symmetry channel
\cite{NatCommun.11.287}.
Finally, composite order parameters
\cite{PhysRevB.87.054503},
excitations of subleading pairing channels
\cite{PhysRevB.100.140501}
as well as coupling to other coexisting modes
\cite{NatCommun.11.1793}
might show up as additional frequencies.

As of now, two main approaches to study Higgs oscillations exist,
namely measurements of the optical conductivity in pump-probe experiments
\cite{PhysRevB.76.224522,PhysRevB.77.180509,PhysRevB.90.014515}
and resonances in third-harmonic generation in periodically driven systems
\cite{PhysRevB.92.064508,PhysRevB.93.180507},
where both methods allow to deduce the intrinsic Higgs modes.
The only different approach, also in a pump-probe setup,
was the prediction of Higgs oscillations in tr-ARPES
\cite{PhysRevB.92.224517,PhysRevB.96.184518,PhysRevB.99.035117},
where the position of the maximum of the energy distribution curve
shows oscillations with the same frequency as the Higgs mode.

In this article, we study the spectral function measured in tr-ARPES
in a more general approach.
While previous papers concentrated on the oscillation
of spectral weight in energy,
we also evaluate the amplitude oscillation,
which contains more information about the condensate dynamic.
We allow arbitrary gap symmetry
and study the effect of quenches in symmetry channels
different from the ground state.
The idea behind such quenches is
to model the net effect of pump pulses in a controlled way,
which act on the condensate momentum-dependently.
These might be realized experimentally
by tuning polarization and pulse direction
\cite{NatCommun.11.287}
or could be implemented in more complex approaches
like transient grating
\cite{Science.300.1410,JChemPhys.120.4755}
or four-wave mixing
\cite{PhysRevLett.109.147403}
setups.

Quenching superconductors with such an approach,
where the symmetry of the condensate is altered
with respect to the ground state,
can result in dynamically created additional Higgs modes
\cite{NatCommun.11.287}.
It is important to note
that we ensure that the order parameter
always keeps its ground state symmetry
and only the underlying condensate is modified.
This is plausible
as a modification of the gap symmetry and thus, the pairing interaction itself,
in a conventional pump-probe experiment is unlikely.
However, a modification of the condensate,
i.e. the Cooper pair or quasiparticle distribution
might be controllable with light excitation.
In our analysis,
we neglect subleading pairing channels,
which may be present in some materials.
A symmetry-breaking quench would activate these channels,
such that Bardasis-Schrieffer modes can occur \cite{PhysRevB.100.140501}.
However, in many materials only a single pairing channel is dominant, such that these effects can be neglected.
As we show in this work,
additional Higgs mode can still occur assuming only a single pairing channel.

By evaluating the induced oscillation of the ARPES intensity,
both in amplitude and weight shift in energy,
we explore the creation process of these additional Higgs modes,
i.e. we can understand
how the dynamic of the condensate at different momenta
contributes to the collective Higgs oscillation.
Moreover,
we compare superconductors
with the same nodal gap structure but different signs,
like $d$-wave and nodal $s$-wave, quenched in the same symmetry channel.
A calculation of the oscillation phase in different lobes
shows opposite phase oscillations,
reflecting the differently induced symmetry deformations of the condensate
in momentum space.

\section{Model and methods}

\subsection{Hamiltonian}
\label{sec:model}
We consider the mean-field BCS Hamiltonian
\begin{align}
    H &= \sum_{\vk\sigma} \epsilon_{\vk}c_{\vk\sigma}^\dagger c_{\vk\sigma}
    - \sum_{\vk} \left(
        \Delta_{\vk} c_{\vk\uparrow}^\dagger c_{-\vk\downarrow}^\dagger
        + \mathrm{c.c.}
    \right)\,,
    \label{eq:bcs_hamiltonian}
\end{align}
where $\epsilon_\vk$ is the electron dispersion relative to the Fermi level
and $c^\dagger_{\vk\sigma}$, $c_{\vk\sigma}$
the creation and annihilation operators for electrons
with momentum $\vk$ and spin $\sigma$.
The momentum-dependent energy gap $\Delta_\vk$ is defined by
\begin{align}
    \Delta_{\vk} &= \sum_{\vk'} V_{\vk\vk'}
    \braket{c_{-\vk'\downarrow}c_{\vk'\uparrow}}\,,
\end{align}
where the pairing interaction
$V_{\vk\vk'} = V f_\vk f_{\vk'}$
is assumed to be separable such that we can write
$\Delta_{\vk} = \Delta f_\vk$ with
\begin{align}
    \Delta = V\sum_{\vk} f_\vk\braket{c_{-\vk\downarrow}c_{\vk\uparrow}}\,.
    \label{eq:gap_eq}
\end{align}
Hereby, $V$ is the pairing strength and $f_\vk$ the gap symmetry function,
which can be chosen according to the symmetry groups of the underlying lattice.
As we will see,
tr-ARPES allows to measure the dynamic of the condensate
$\braket{c_{-\vk\downarrow}c_{\vk\uparrow}}(t)$
by measuring the spectral function,
such that we can understand how oscillations of
$\braket{c_{-\vk\downarrow}c_{\vk\uparrow}}(t)$ at different momenta
contribute to collective Higgs oscillation $\Delta(t)$.

Our results are rather independent of material parameters
(see appendix \ref{sec:influence_parameters}),
thus, we use a quadratic dispersion
$\epsilon_\vk = t k^2 - \epsilon_\mathrm F$ for convenience.
The used numerical values of the parameters can be found in the appendix
in Tab.~\ref{tab:params}.
Furthermore, we restrict our calculation to two dimensions
to reduce the computational costs
and to correspond also to quasi-2d materials as the layered cuprates.

To calculate the induced time evolution after a quench,
it is advantageous to perform a Bogoliubov transformation
\begin{align}
    \VV{\alpha_{\vk}}{\beta^\dagger_{\vk}}
    &= \MM{u_{\vk}}{-v_{\vk}}{v_{\vk}}{u_{\vk}}
    \VV{c_{\vk\uparrow}}{c_{-\vk\downarrow}^\dagger}
    \label{eq:bt_transform}
\end{align}
to new quasiparticle operators $\alpha_\vk$ and $\beta_\vk$ with
\begin{align}
    u_{\vk} &=
        \sqrt{\frac 1 2\left(1+\frac{\epsilon_{\vk}}{E_{\vk}}\right)}\,,&
    v_{\vk} &=
        \sqrt{\frac 1 2\left(1-\frac{\epsilon_{\vk}}{E_{\vk}}\right)}\,.
\end{align}
In the expression above,
$E_\vk = \sqrt{\epsilon_\vk^2 + (\Delta f_\vk)^2}$ is the quasiparticle energy
and we choose the phase of the equilibrium order parameter $\Delta$ to be zero.
The Bogoliubov transformation is performed for this fixed value of the gap.
The BCS Hamiltonian at arbitrary times
for time-dependent values of the order parameter $\Delta(t)$
in Eq.~\eqref{eq:bcs_hamiltonian} then reads
\begin{align}
    H(t) &= \sum_{\vk} R_{\vk}(t)(
    \alpha_{\vk}^\dagger\alpha_{\vk}
    + \beta_{\vk}^\dagger\beta_{\vk}
    )
    \notag\\&\quad
    + \sum_{\vk} C_{\vk}(t) \alpha_{\vk}^\dagger\beta_{\vk}^\dagger
    - C_{\vk}^*(t) \alpha_{\vk}\beta_{\vk}\,,
\end{align}
where we neglect constant terms and define
\begin{subequations}
\begin{align}
    R_{\vk}(t) &= \frac{1}{2E_{\vk}}\left(
    2\epsilon_{\vk}^2 + \Delta_{\vk}(t)\Delta_\vk^*
    + \Delta_{\vk}^*(t)\Delta_\vk
    \right)\,,\\
    C_{\vk}(t) &= \frac{1}{2E_{\vk}}\big(2\epsilon_{\vk}\Delta_\vk
    - \Delta_{\vk}(t)(E_{\vk} + \epsilon_{\vk})
    \notag\\&\qquad
    + \Delta_{\vk}^*(t)(E_{\vk} - \epsilon_{\vk})\big)\,.
\end{align}
\end{subequations}
In equilibrium, $\Delta_{\vk}(t) = \Delta_\vk$
and the Hamiltonian becomes diagonal with $R_\vk = E_\vk$ and $C_\vk = 0$.
We can express the anomalous expectation value
$\braket{c_{-\vk\downarrow}c_{\vk\uparrow}}$
in the gap equation \eqref{eq:gap_eq}
with the new Bogoliubov quasiparticle operators and find
\begin{align}
    \Delta(t) &= V\sum_{\vk} f_\vk \Big[
    u_{\vk}v_{\vk}\left(
        1 - \braket{\aDa}(t)
        - \braket{\bDb}(t)\right)
    \notag\\&\qquad
    - u_{\vk}^2\braket{\ab}(t)
    - v_{\vk}^2\braket{\aDbD}(t)
    \Big]\,.
    \label{eq:gap_equation_qp}
\end{align}

\subsection{Iterated equation of motion method}
We calculate the time evolution of the Bogoliubov operators
with the help of Heisenberg's equation of motion
\begin{subequations}
\begin{align}
    \partial_t \alpha_{\vk}^\dagger(t) &=
        \frac{\im}{\hbar} [\tilde H(t), \alpha_{\vk}^\dagger(t)]\,,
    \\
    \partial_t \beta_{\vk}^\dagger(t) &=
        \frac{\im}{\hbar} [\tilde H(t), \beta_{\vk}^\dagger(t)]\,,
\end{align}
\label{eq:heom}%
\end{subequations}
where the Hamiltonian $\tilde H(t)$,
$\alpha_{\vk}^\dagger(t)$ and $\beta_{\vk}^\dagger(t)$
are the operators in the Heisenberg picture.
To evaluate these equations,
we make use of the iterated equation of motion approach
\cite{EurPhysJB.90.97}
to write the Bogoliubov operators in the Heisenberg picture
as a time-dependent superposition of the equilibrium operators
\begin{subequations}
\begin{align}
    \alpha_\vk^\dagger(t) &= a_{0\vk}(t)\alpha_\vk^\dagger
        + a_{1\vk}(t)\beta_\vk\,, \\
    \beta_\vk^\dagger(t) &= b_{0\vk}(t)\beta_\vk^\dagger
        + b_{1\vk}(t)\alpha_\vk\,.
\end{align}
\label{eq:ieom_ansatz}%
\end{subequations}
Inserted into Heisenberg's equation of motion,
such an ansatz leads in general to an infinite hierarchy of equations
for the prefactors $a_{i\vk}(t)$ and $b_{i\vk}(t)$,
which have to be truncated at a certain order.
However, for a bilinear Hamiltonian like the BCS Hamiltonian,
the approach in Eq.~\eqref{eq:ieom_ansatz}
using only a second order ansatz for each operator is exact.
This allows to derive a closed set of differential equations
for the time-dependent prefactors.
The derived equations are given in appendix~\ref{sec:eom}.

The coupled differential equations \eqref{eq:eom_prefactors}
are solved numerically in a self-consistent manner
by evaluating the time-dependent gap equation in each time step.
The initial values of the prefactors $a_{i\vk}(0)$ and $b_{i\vk}(0)$
are given by the initial values
of the Bogoliubov quasiparticles expectation values
$\braket{\alpha_\vk^\dagger\alpha_\vk}(0)$,
$\braket{\beta_\vk^\dagger\beta_\vk}(0)$,
$\braket{\alpha_\vk^\dagger\beta_\vk^\dagger}(0)$
and $\braket{\alpha_\vk\beta_\vk}(0)$,
which themselves are determined by the electron expectation values
$\braket{c_{-\vk\downarrow}c_{\vk\uparrow}}(0)$,
$\braket{c_{\vk\uparrow}^\dagger c_{-\vk\downarrow}^\dagger}(0)$,
$\braket{c_{\vk\uparrow}^\dagger c_{\vk\uparrow}}(0)$
and $\braket{c_{-\vk\downarrow} c_{-\vk\downarrow}^\dagger}(0)$.
Depending on the type of quench,
the initial values are either the equilibrium expectation values
or quenched expectation values.
This is described in the next section
and the respective expectation values are listed in appendix~\ref{sec:eom}.

In order to increase the precision
while keeping the numerical effort in a reasonable order,
the 2d momentum space is discretized only in a small region
around the Fermi level given by an energy cutoff $\epsilon_\mathrm c$.
This is justified by the fact
that superconductivity only occurs near the Fermi momentum $k_\mathrm F$,
while contributions far away are negligible.
The momentum grid was chosen in polar coordinates,
with $N_k$ points in radial and $N_\varphi$ points in azimuthal direction.
All the numerical values are listed in the appendix in Tab.~\ref{tab:params}.

\subsection{Excitation with quantum quenches}
\label{sec:quenches}
We want to study the system in an out of equilibrium state
excited by a laser pulse in an experiment.
As the net effect of an ultrashort THz pump pulse
is that it changes the system abruptly,
we model this effect in a general manner
by considering an effective quantum quench,
i.e. the pump pulse acts like a quench pulse.
In literature
\cite{PhysRevLett.96.230404},
the usual way to go is an \emph{interaction quench},
i.e. changing the interaction strength abruptly at $t=0$
from its equilibrium value $V$ to a new value $V' = g V$,
where $g < 1$,
which reduces the gap $\Delta$ to a new value $\Delta(0) = g\Delta$.
Then, the time evolution can be calculated
starting from the equilibrium state of the old system
given in Eq.~\eqref{eq:ai0bi0_eq}
using the quenched gap equation with the modified pairing interaction $V'$.
The disadvantage of this method is that an interaction quench
cannot easily be realized experimentally in condensed matter.
Only in cold atom systems, where the interaction strength
can be tuned with Feshbach resonances
\cite{RevModPhys.82.1225},
an implementation is feasible.
Besides that, an interaction quench is not directly equivalent
to the effect of a quench pulse
and in addition, it always acts on the whole system isotropically,
whereas a momentum-dependent excitation is more interesting to study,
especially for unconventional superconductors
with nontrivial pairing symmetry.

Considering all of this, we implement a momentum-dependent \emph{state quench}.
Namely, the effect of a quench pulse
is to deplete a small portion of the condensate
and create a nonequilibrium quasiparticle distribution.
What is happening in particular,
is that the equilibrium distribution of the quasiparticles at $T = 0$,
e.g. the anomalous expectation value
\begin{align}
    \braket{c_{-\vk\downarrow}c_{\vk\uparrow}} &= \frac{\Delta_\vk}{2E_\vk}\,,
\end{align}
is modified as a result of the quench pulse,
where the symmetry of the quenched state
$\braket{c_{-\vk\downarrow}c_{\vk\uparrow}}'$
does not necessarily have to be the same as in equilibrium.
To implement such a quench in a controlled way,
we introduce the quenched distribution
\begin{align}
    \braket{c_{-\vk\downarrow}c_{\vk\uparrow}}'
        &= \frac{\Delta_\vk'}{2E_\vk'}\,,
    \label{eq:state_quench}
\end{align}
which is an equilibrium distribution for a different symmetry $f_\vk'$
with $\Delta_\vk' = \Delta f_\vk'$
and $E_\vk' = \sqrt{\epsilon_\vk^2 + (\Delta f_\vk')^2}$.
This distribution is of course no longer the equilibrium distribution
for the original system and the deviation can be controlled
by the modified symmetry function
\begin{align}
    f_\vk' = f_\vk + g f_\vk^{\mathrm q},
\end{align}
which is a small deviation of strength $g$ from the original symmetry $f_\vk$
with the quench symmetry $f_\vk^{\mathrm q}$.
In addition to the exemplary shown anomalous expectation value,
all other expectation values are modified with the same $f_\vk'$ as well.
This is shown in appendix~\ref{sec:eom}.
Using the quenched expectation values of the quasiparticles,
the quenched prefactors $a_{i\vk}(0)$ and $b_{i\vk}(0)$ can be calculated
according to Eq.~\eqref{eq:ai0bi0}.
With these initial values, Heisenberg's equation of motion can be integrated.

\subsection{ARPES}
The time-dependent ARPES intensity $A(\vk,\omega,t)$ can be calculated
with the help of the lesser Green's function $G^<(\vk,\omega,t)$
\cite{PhysRevLett.102.136401,PhysRevB.92.224517,PhysRevB.99.035117}
\begin{align}
    A(\vk,\omega,t) &= \Im \int\,\mathrm dt_2 \, \int \mathrm dt_1\,
    \Big( G^<(\vk,t_1,t_2)
    \notag\\&\qquad\times
        p(t_1-t) p(t_2-t) \e^{\im\omega(t_1-t_2)} \Big)\,.
    \label{eq:arpes_intensity}
\end{align}
Hereby, we neglect any matrix element effects.
The lesser Green's function in the time-domain, is defined as
\begin{align}
    G^<(\vk,t_1,t_2) &= \im
    \braket{c_{\vk\uparrow}^\dagger(t_2) c_{\vk\uparrow}(t_1)}\,.
    \label{eq:gf_time}
\end{align}
The required electron expectation value at different times
can be computed from the time evolution
of the Bogoliubov quasiparticle operators
\begin{align}
    &\braket{c_{\vk\uparrow}^\dagger(t_2)c_{\vk\uparrow}(t_1)} =
    u_\vk^2 \braket{\alpha_\vk^\dagger(t_2)\alpha_\vk(t_1)}
    \notag\\&\quad
        + v_\vk^2 \Big(
        b_{0\vk}^*(t_2)b_{0\vk}(t_1) + b_{1\vk}^*(t_2)b_{1\vk}(t_1)
        - \braket{\beta_\vk^\dagger(t_1)\beta_\vk(t_2)}
        \Big)
    \notag\\&\quad
        + u_{\vk} v_\vk \Big(
            a_{0\vk}^*(t_1)b_{1\vk}^*(t_2) + a_{1\vk}^*(t_1)b_{0\vk}^*(t_2)
    \notag\\&\qquad
            - \braket{\alpha_\vk(t_1)\beta_\vk(t_2)}
        \Big)
        + u_\vk v_{\vk} \braket{\alpha_\vk^\dagger(t_2)\beta_\vk^\dagger(t_1)}\,.
\end{align}
The finite width of the probe pulse broadens the spectral function.
Thus, in the expression for the ARPES intensity Eq.~\eqref{eq:arpes_intensity},
the probe pulse envelope is incorporated by a Gaussian function
\begin{align}
    p(t) = \sqrt{\frac{4\log 2}{\pi \tau_p^2}}
        \exp\left(-4\log 2 \left(\frac{t}{\tau_p}\right)^2\right)
\end{align}
with full width at half maximum $\tau_p$.

\section{Results}
In the following,
we will investigate the dynamics of two quantities.
On the one hand,
we will consider the position of the maximum
of the energy distribution curve (EDC) at $k=k_\mathrm F$, i.e.
\begin{align}
    \mathcal E(t,\varphi) &=
        \operatorname{argmax}_\omega A(k=k_\mathrm F,\varphi,\omega,t)\,.
    \label{eq:EDCmax}
\end{align}
This quantity, i.e. the position of the maximum with respect to the Fermi level
reflects the energy gap and thus,
directly reveals any collective excitation of the order parameter.
On the other hand,
we will follow the dynamic of the amplitude of the ARPES intensity
at $\omega = -\Delta_\infty$,
where the strongest dynamic can be expected, i.e.
\begin{align}
    \mathcal A(t,k,\varphi) &= A(k,\varphi,\omega=-\Delta_\infty,t)\,.
    \label{eq:A}
\end{align}
In these expressions,
the momentum is expressed in polar coordinates
with absolute value $k$ and polar angle $\varphi$
and $\Delta_\infty$ is the value of the gap after a long time.
While the first quantity is typically used
to trace the time evolution of the system
\cite{PhysRevB.92.224517,PhysRevB.96.184518},
it will appear that an investigation of the second quantity
yields even more information.
In contrast to the first quantity,
it does not reveal directly the collective modes of the system
but the dynamic of the underlying condensate and allows to understand
how it contributes and creates
the collective excitation of the order parameter.

\subsection{Interaction quench for $s$-wave superconductor}
\begin{figure}[t]
    \centering
    \includegraphics[width=\columnwidth]{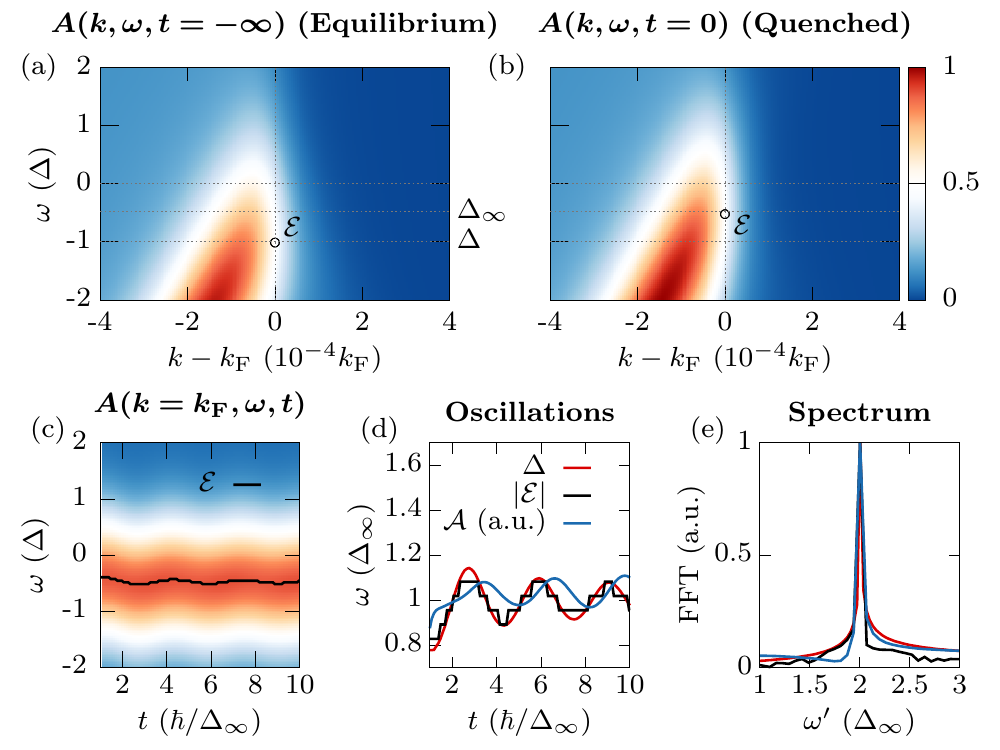}
    \caption{\label{fig:swave_time_evol}%
    ARPES intensity of an $s$-wave superconductor
    after performing an interaction quench.
    (a) Equilibrium ARPES intensity $A(k,w,t=-\infty)$,
    the maximum of the energy distribution curve (EDC)
    at $k=k_\mathrm F$ is marked by a circle and labeled $\mathcal E$.
    The maximum is at $|\mathcal E| = \Delta$.
    (b) ARPES intensity $A(k,w,t=0)$ after the interaction quench,
    spectral weight is shifted towards the Fermi level.
    The maximum is at the reduced value $\Delta_\infty < |\mathcal E| < \Delta$.
    (c) Time evolution of the EDC at $k=k_\mathrm F$.
    See also Fig.~\ref{fig:energ_vs_ampl}.
    (d) Gap oscillations $\Delta(t)$,
    oscillations of the EDC maximum $|\mathcal E(t)|$
    and amplitude oscillation $\mathcal A(t,k=k_\mathrm F)$
    of the ARPES intensity.
    The edges in $\mathcal E(t)$
    are an artifact due to the finite frequency resolution.
    (e) Fourier transform of oscillations,
    all quantities oscillate with the Higgs frequency $\omega=2\Delta_\infty$.
    The probe pulse width is $\tau_p = 1$\,$\hbar/\Delta_\infty$.}
\end{figure}

\begin{figure}[t]
    \centering
    \includegraphics[width=\columnwidth]{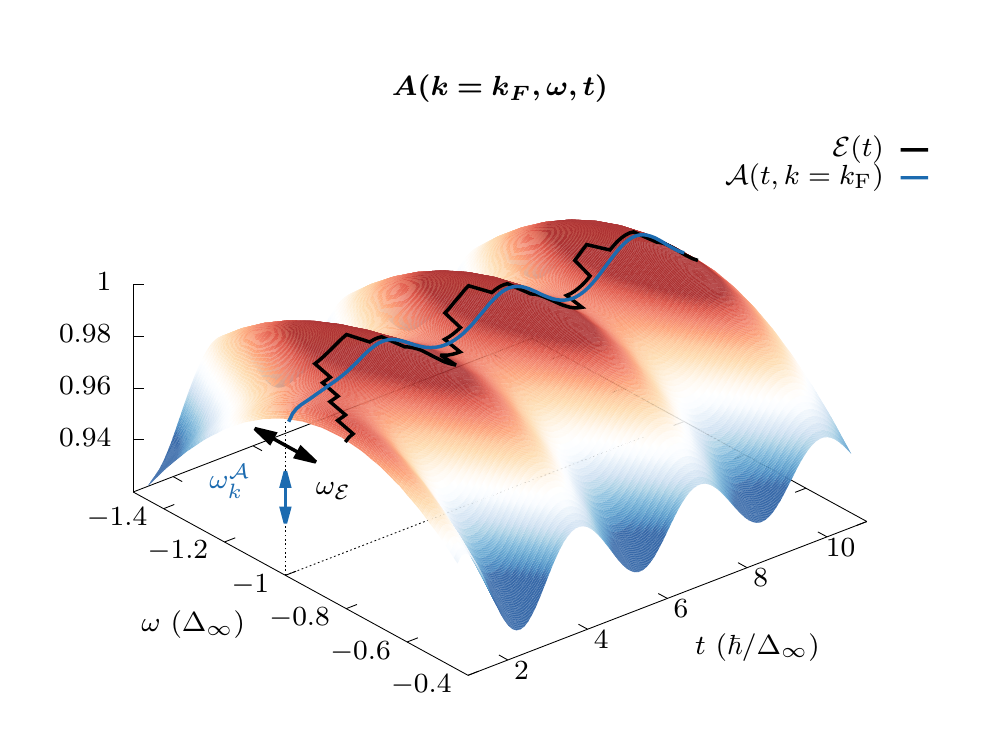}
    \caption{\label{fig:energ_vs_ampl}%
    Time evolution of the EDC at $k=k_\mathrm F$.
    It corresponds to Fig.~\ref{fig:swave_time_evol}(c) plotted
    as a 3d plot for better visualization.
    There are two different kinds of dynamics:
    Oscillations in energy of the maximum of the EDC curve (black line)
    as defined in Eq.~\eqref{eq:EDCmax}
    with the momentum-independent frequency
    $\omega_\mathcal E = 2\Delta_\infty$
    and oscillations in the amplitude of the ARPES intensity (blue line)
    as defined in Eq.~\eqref{eq:A}
    with the momentum-dependent frequency
    $\omega_\vk^\mathcal A = 2\sqrt{\epsilon_\vk^2+(\Delta_\infty f_\vk)^2}$.
    At $k=k_\mathrm F$ and $f_\vk = 1$,
    $\omega_\vk^\mathcal A = \omega_\mathcal E = 2\Delta_\infty$.}
\end{figure}

Before studying superconductors with nontrivial gap symmetry,
we first consider the $s$-wave case with $f_\vk = 1$,
where we perform a simple interaction quench to trigger the time dynamic.
In Fig.~\ref{fig:swave_time_evol},
the time evolution of the ARPES intensity can be found.
One observes that after the quench (Fig.~\ref{fig:swave_time_evol}(b)),
spectral weight is shifted towards and above the Fermi level
compared to the equilibrium case (Fig.~\ref{fig:swave_time_evol}(a)).
This can especially be seen by determining $\mathcal E$
as defined in Eq.~\eqref{eq:EDCmax}.
This position of the maximum,
indicated by a circle in Fig.~\ref{fig:swave_time_evol}(a)
and Fig.~\ref{fig:swave_time_evol}(b)
is shifted from its initial value $|\mathcal E(t=-\infty)| = \Delta$
to a smaller value $|\mathcal E(t=0)| < \Delta$,
which indicates a suppression of the energy gap after the quench.
To trace the induced dynamic,
we look at the time evolution of the EDC at $k=k_\mathrm F$ shown
in Fig.~\ref{fig:swave_time_evol}(c) and Fig.~\ref{fig:energ_vs_ampl},
where we extract the two quantities
$\mathcal E(t)$ and $\mathcal A(t,k=k_\mathrm F)$.

For a sufficiently short probe pulse,
i.e. $\tau_p$ is small compared
to the timescale given by $\hbar/(2\Delta_\infty)$,
the time-resolution is high enough to directly observe
the induced Higgs oscillations of the order parameter
as oscillations of $\mathcal E(t)$,
i.e. an oscillation of the weight of the ARPES intensity in energy.
The extracted curve $\mathcal E(t)$
is shown in Fig.~\ref{fig:swave_time_evol}(d)
in comparison to the calculated dynamic of the order parameter $\Delta(t)$,
where a close accordance can be observed.
The oscillation starts at the quenched value $\Delta(0) < \Delta$
and oscillates around $\Delta_\infty < \Delta(0)$ in the long-time limit.

In addition to the oscillation of $\mathcal E(t)$,
we also look at the dynamic
of the amplitude of the ARPES intensity $\mathcal A(t, k=k_\mathrm F)$
as defined in Eq.~\eqref{eq:A},
shown as well in Fig.~\ref{fig:swave_time_evol}(d) for comparison.
We find oscillations with the same frequency as the order parameter.
This is confirmed in Fig.~\ref{fig:swave_time_evol}(e),
where the Fourier transform of the order parameter $\Delta(t)$,
the maximum of the EDC $\mathcal E(t)$
and the ARPES intensity $\mathcal A(t,k=k_\mathrm F)$ are compared.
All three quantities show the same oscillation
with a frequency of $\omega = 2\Delta_\infty$,
the energy of the Higgs mode.

The pulse width plays a crucial role
for the resolution of the gap dynamic in $\mathcal E$ and $\mathcal A$.
If the probe pulse is too wide, i.e. $\tau_p \gg \hbar/(2\Delta_\infty)$,
it cannot resolve the oscillations of the system
and $\mathcal E$ and $\mathcal A$
will be constant in time at their mean value,
which is $\Delta_\infty$ in the case of $\mathcal E$,
i.e. the value of the order parameter after a long time.
The influence of the probe pulse width is discussed in more detail
in appendix~\ref{sec:pulse_width}.

Thus, the ARPES intensity provides a direct measure
of Higgs oscillations for superconductors in nonequilibrium,
both in energy and amplitude.

\subsection{State quench for $d$-wave superconductor}
\label{sec:quench_dwave}
\begin{figure}[t]
    \centering
    \includegraphics[width=\columnwidth]{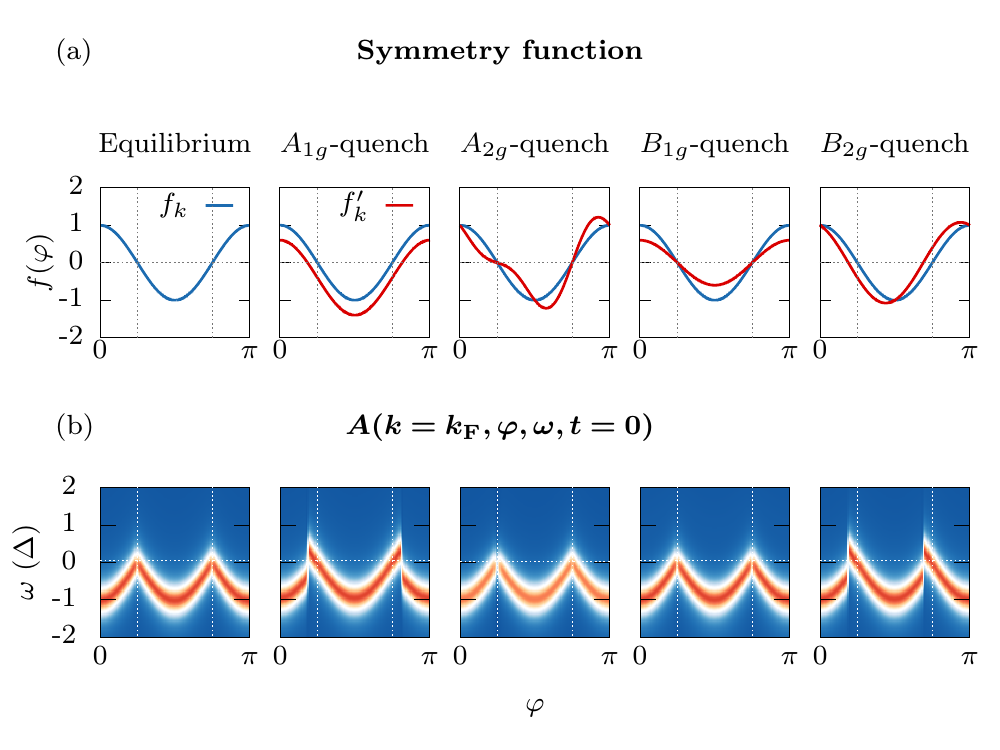}
    \caption{\label{fig:dwave_quench}%
    ARPES intensity of a $d$-wave superconductor after a state quench.
    (a) Symmetry function $f_\vk$ (blue)
    and quenched symmetry function $f_\vk'$ (red) for different channels.
    (b) Corresponding ARPES intensity.
    A shift of nodal lines is reflected by a shift of spectral weight
    in the nodal region.
    The probe pulse width is $\tau_p = 10$\,$\hbar/\Delta_\infty$.}
\end{figure}

After having considered the $s$-wave case with an isotropic interaction quench,
we will proceed to a $d$-wave superconductor,
which we will quench in different symmetry channels
with the help of a state quench as defined in Sec.~\ref{sec:quenches}.
The symmetry function of a $d$-wave superconductor close to the Fermi level
can be written as $f_\vk = \cos(2\varphi)$, where $\varphi$ is the polar angle.
In order to excite the system in a systematic way,
we choose quench symmetries
which are fundamental with respect to the lattice point group.
To this end, the superconductor is quenched in all different symmetry channels
for the $D_{4h}$ point group,
which is the underlying lattice point group
for the important $d$-wave example of cuprate superconductors.
We define the following four quenches:
\begin{subequations}
\begin{align}
    A_{1g}:&\quad f_\vk^\mathrm q = 1\,,\\
    A_{2g}:&\quad f_\vk^\mathrm q = \sin(4\varphi)\,,\\
    B_{1g}:&\quad f_\vk^\mathrm q = \cos(2\varphi)\,,\\
    B_{2g}:&\quad f_\vk^\mathrm q = \sin(2\varphi)\,.
\end{align}
\label{eq:quenches}%
\end{subequations}
The symmetry function and the quenched symmetry function,
as well as the ARPES intensity after the quench,
can be found in Fig.~\ref{fig:dwave_quench}.
First of all, in the equilibrium case,
the ARPES intensity closely resembles
the absolute value of the symmetry function.
In the antinodal directions at $\varphi = [0,\pi/2,\pi,3\pi/2]$,
a full open gap can be observed with $|\mathcal E| = \Delta$.
Moving $\varphi$ towards the nodal directions
$\varphi = [\pi/4,3\pi/4,5\pi/4,7\pi/4]$,
the maximum $\mathcal E$ approaches zero.

The influence of the different quenches is also reflected
in the change in the ARPES intensity.
The $A_{1g}$- and $B_{2g}$-quenches shift the position of the nodes.
In the case of the $A_{1g}$-quench,
the shift is in opposite directions relative to the lobe maxima,
which creates lobes with different sizes.
The $B_{2g}$-quench shifts all nodes in the same direction,
which results in a rotation of all lobes.
This can be seen by a transfer of spectral weight above the Fermi level
at exactly these points in the ARPES intensity.
The $B_{1g}$-quench does not change the symmetry at all
and the $A_{2g}$-quench only shifts weight inside the lobes,
such that for these two quenches, no obvious change
in the symmetry of the ARPES intensity is observable.

\begin{figure}[t]
    \centering
    \includegraphics[width=\columnwidth]{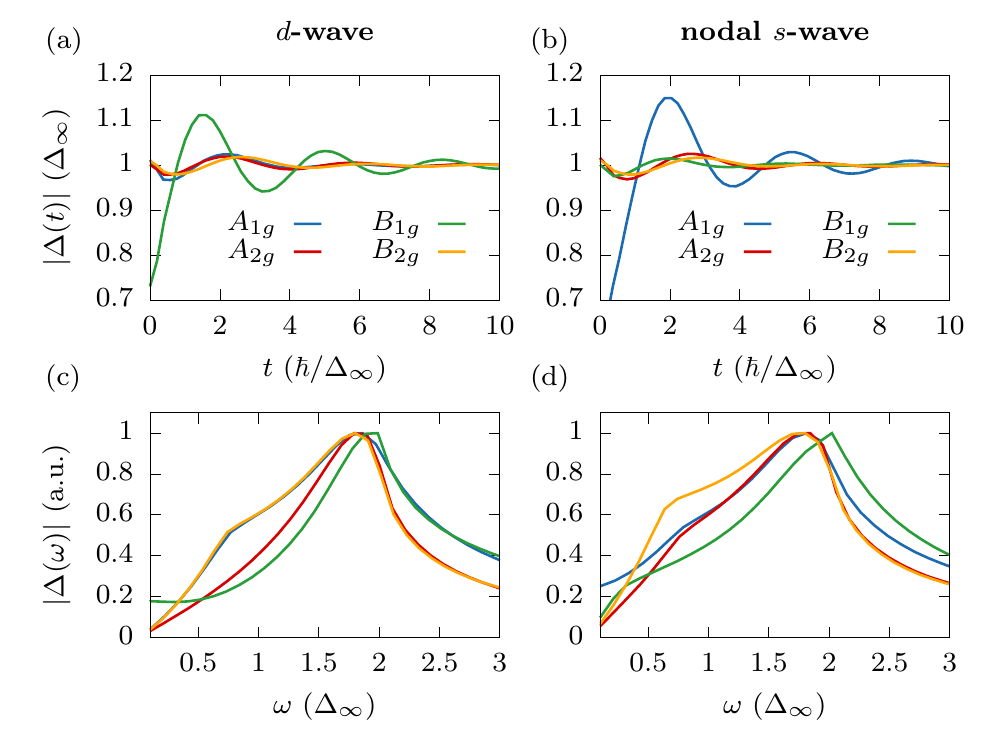}
    \caption{\label{fig:gap_osci}%
    Gap dynamic of $d$-wave (left) and nodal $s$-wave (right).
    Top: Oscillations after quenching in the four fundamental symmetries
    defined in Eq.~\eqref{eq:quenches}.
    The $B_{1g}$-quench for $d$-wave and the $A_{1g}$-quench for nodal $s$-wave
    induce stronger oscillations
    as these quenches are within the same symmetry channel
    as the gap symmetry.
    Bottom: Fourier transform of the gap oscillations.
    All energies are normalized to their individual $\Delta_\infty$.}
\end{figure}

This fact is crucial for the induced Higgs oscillations of the gap,
as for the $A_{1g}$- and $B_{2g}$-quenches,
the shift of the nodal lines creates a second low-lying Higgs mode dynamically,
whereas for the other two quenches, it will not.
This can be seen in Fig.~\ref{fig:gap_osci}(a) and (c),
where the oscillations of the order parameter and their Fourier transform
is shown for the different quench symmetries.
While the $B_{1g}$- and $A_{2g}$-quenches
only show a single broad peak around $\omega = 2\Delta_\infty$,
the $A_{1g}$- and $B_{2g}$-quenches
show a two peak or kink structure in the spectrum
with one peak around $\omega = 2\Delta_\infty$
and a second peak at lower frequencies.
The same behavior is visible for nodal $s$-wave,
which is discussed in appendix~\ref{sec:nod_s}
and shown in Fig.~\ref{fig:gap_osci}(b) and (d).
The usual $2\Delta$ Higgs mode
corresponds to an oscillation in the ground state symmetry channel,
whereas the low-lying Higgs mode
corresponds to an oscillation in another symmetry channel
\cite{NatCommun.11.287}.
The energies of these two modes are controlled by different quantities.
While the energy of the $2\Delta$ Higgs mode is given by the value of the gap,
i.e. $\Delta_\infty$ after the quench,
the energy of the other mode is controlled
by the strength of deviation from the ground state symmetry.
Thus, for increasing quench strength,
the energy of the $2\Delta$ Higgs mode decreases,
while the energy of the low-lying Higgs mode increases.
This is discussed in appendix~\ref{sec:mdep_freq} in more detail.

Now, we will analyze the amplitude oscillations
of the ARPES intensity $\mathcal A(t)$ in more detail
to gain a deeper understanding of the dynamic creation
of the low-lying Higgs mode.
According to the gap equation~\eqref{eq:gap_eq},
the value of $\Delta(t)$ is obtained
by a summation over the condensate
$\braket{c_{-\vk\downarrow}c_{\vk\uparrow}}$(t).
Both, the anomalous Green's function
$\braket{c_{-\vk\downarrow}c_{\vk\uparrow}}$(t)
and the normal Green's function
$\braket{c_{\vk\uparrow}^\dagger c_{\vk\uparrow}}$(t)
share a similar dynamic as their equations of motion are coupled.
This can be seen in Eq.~\eqref{eq:bloch_linear}
in appendix \ref{sec:mdep_freq}.
Therefore, a momentum-resolved analysis of the ARPES intensity,
i.e. $\mathcal A(t,k,\varphi)$,
which is proportional to the normal Green's function,
can reveal important information about the dynamic processes.
In comparison,
an angle-resolved evaluation of the maximum of the EDC curve, i.e.
$\mathcal E(t,\varphi)$,
as considered in \cite{PhysRevB.96.184518},
does not give further insight
as it traces only the momentum-averaged quantity $\Delta(t)$.

\begin{figure}[t]
    \centering
    \includegraphics[width=\columnwidth]{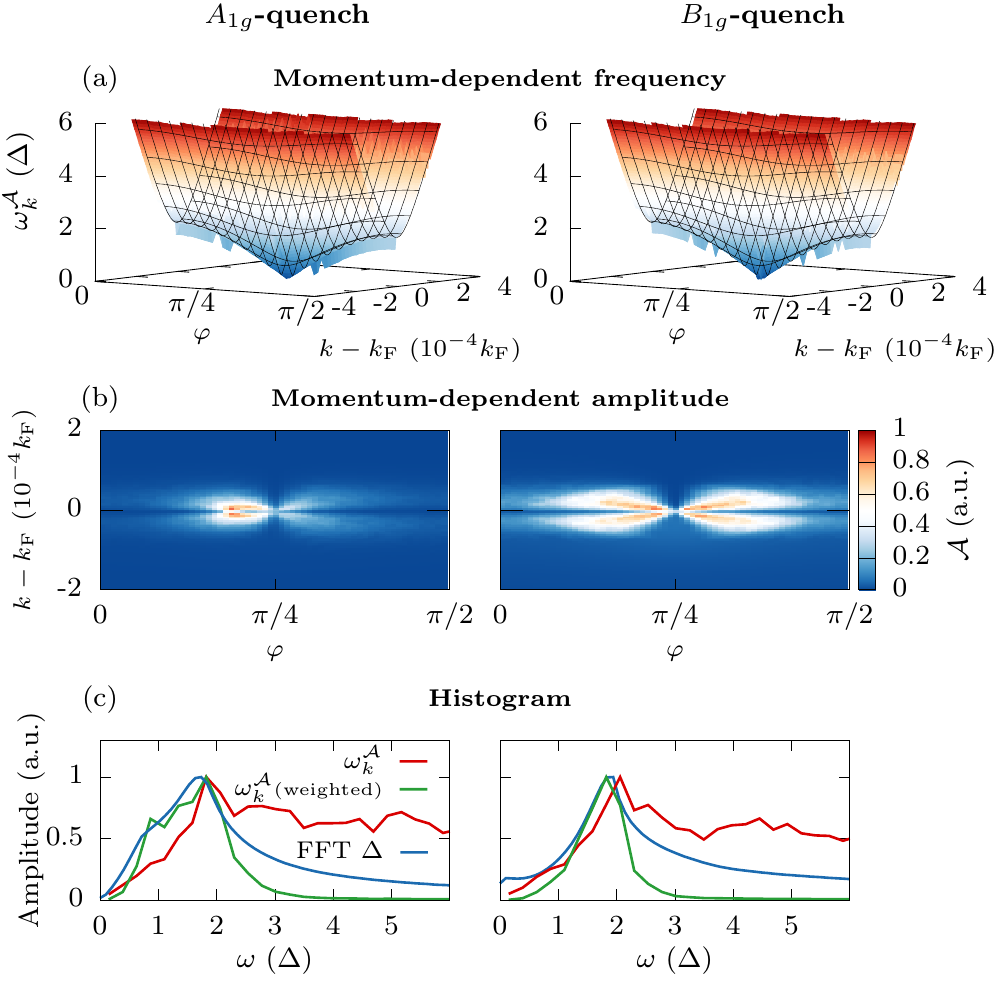}
    \caption{\label{fig:freq_histogram}%
    Frequency distribution of ARPES intensity oscillation
    for $A_{1g}$-quench (left) and $B_{1g}$-quench (right).
    (a) Momentum-dependent frequency $\omega_\vk^{\mathcal A}$
    of $\mathcal A(t,k,\varphi)$ (colored)
    compared with the theoretical formula from Eq.~\eqref{eq:winf} (lines).
    (b) Momentum-dependent amplitudes of the oscillations.
    (c) Histogram of occurring frequencies
    compared with Fourier transform of gap oscillations.
    The probe pulse width is $\tau_p = 1$\,$\hbar/\Delta_\infty$.}
\end{figure}

To this end, we extract $\mathcal A(t,k,\varphi)$
and its frequency $\omega_{\vk}^{\mathcal A}$
not only at $k=k_{\mathrm F}$ but for all points in momentum space
and plot its distribution in Fig.~\ref{fig:freq_histogram}(a)
for a $d$-wave superconductor quenched in the $A_{1g}$- and $B_{1g}$-channels.
The resulting frequencies are momentum-dependent
yet do not depend on the quench.
Namely, the frequency at each momentum point $\vk$ is given by
two times the quasiparticle energy
\begin{align}
    \omega_{\vk}^{\mathcal A} =
        2\sqrt{\epsilon_\vk^2 + (\Delta_\infty f_\vk)^2}\,.
    \label{eq:winf}
\end{align}
This result is derived in appendix~\ref{sec:mdep_freq}.
The same result for $s$-wave symmetry and an interaction quench
was also found in \cite{PhysRevLett.96.230404}.
A comparison of the extracted frequencies
and the analytic formula shows perfect agreement.

In the summation process of the gap equation,
the frequency with the largest weight will dominate the oscillations.
In the case of $s$-wave with $f_\vk=1$, this turns out to be at $k=k_F$,
where $\epsilon_\vk = 0$,
which results in a single main frequency $\omega = 2\Delta_\infty$.
However, as the other frequencies still contribute,
the superposition of all different frequencies
drives the oscillations out of phase in the long-time limit,
which results in the well-known $1/\sqrt{t}$ decay
\cite{SovPhysJETP.38.1018}.
In the case of $d$-wave, there is a much larger variety of frequencies
due to the angular dependence of the gap $\Delta_\vk$.
Therefore, the summation process creates a much stronger decay
\cite{PhysRevLett.115.257001}.
As the oscillations in $\mathcal E(t)$
correspond directly to the gap oscillations $\Delta(t)$,
the damping due to the dephasing is also visible in this quantity.
In contrast, the oscillations visible in $\mathcal A(t,k,\varphi)$
are undamped as they correspond to oscillations of the quasiparticles,
which can be understood as a precession of pseudospins
as shown in appendix~\ref{sec:mdep_freq}.
In the gap equation, it is exactly this undamped oscillation
of different momentum-dependent frequencies which is summed up
leading to the dephasing effect
and the damping of the collective Higgs oscillations.
The analytic expression for the pseudospin precession
derived in Eq.~\eqref{eq:dxdz} in the appendix shows this undamped oscillation.

The second low-lying frequency for certain quenches
can be explained in the same picture.
In Fig.~\ref{fig:freq_histogram}(b),
we show the extracted amplitudes
for $\omega_{\vk}^{\mathcal A}$ at each momentum point,
which correspond to the actual weight for each frequency.
While there is no difference in the frequency distribution
between different quenches in Fig.~\ref{fig:freq_histogram}(a),
we can clearly see a difference in the amplitudes between the $A_{1g}$-quench,
which shows a second peak in the Higgs oscillations,
and the $B_{1g}$-quench, which shows no second peak.
The change in symmetry due to the quench creates a strong asymmetry
in the weights regarding the positive and negative lobe direction
in the case of the $A_{1g}$-quench,
while the amplitudes for the $B_{1g}$-quench are symmetric.
This can result in the summation process in an additional enhancement
of frequencies other than $\omega = 2\Delta_\infty$.
In Fig.~\ref{fig:freq_histogram}(c),
we show the histogram of the frequencies from all momentum points.
The raw histogram (red curve) looks exactly the same for both quenches,
with a peak at $\omega = 2\Delta_\infty$ already without additional weighting.
Now, if we weight the histogram with the amplitude of the oscillation,
the picture changes (green curve).
In the case of the $A_{1g}$-quench, a two peak or two kink structure
becomes visible, very similar to the Fourier transform
of the gap oscillation (blue curve),
whereas the symmetric weighting in the case of the $B_{1g}$-quench
only shows a one peak structure.
Hence, a careful analysis of the momentum-resolved frequencies
in the ARPES spectrum reveals the dynamic of the condensate
and explains the underlying processes
leading to the collective Higgs oscillations.

\subsection{Analysis of oscillation phase}
\begin{figure}[t]
    \centering
    \includegraphics[width=\columnwidth]{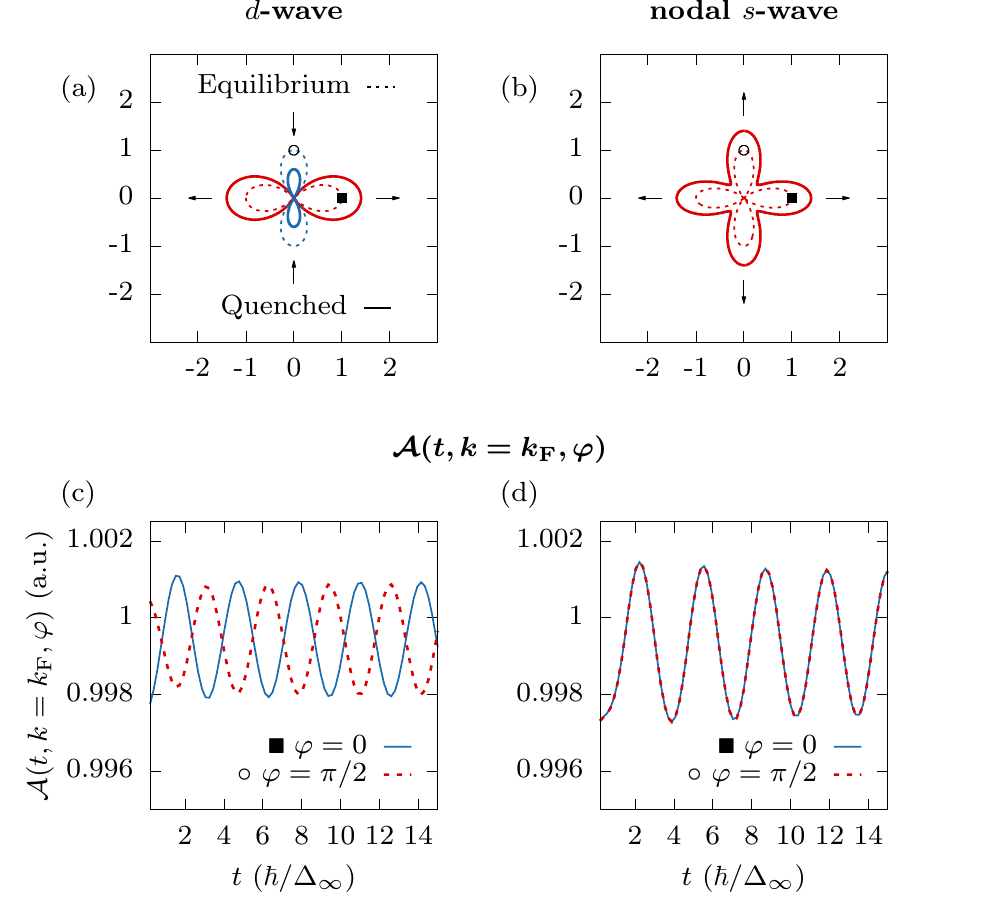}
    \caption{\label{fig:phase_analysis}%
    Oscillation of ARPES intensity after $A_{1g}$-quench for $d$-wave (left)
    and nodal $s$-wave (right) superconductor.
    (a),(b) Schematic picture of condensate symmetry in equilibrium (dotted)
    and quenched (solid lines).
    The arrows indicate the movement of the lobes.
    Points in momentum space rotated by $\pi/2$
    oscillate with opposite phase in the case of $d$-wave
    and in phase in the case of nodal $s$-wave.
    For a $B_{1g}$-quench, the situation would be reversed.
    (c),(d) Time-dependent ARPES intensity.
    The oscillation of the ARPES intensity
    follows the oscillation of the lobes.
    The out of phase oscillation of $d$-wave lobes is clearly visible,
    in comparison to the in phase oscillations of the nodal $s$-wave.
    The probe pulse width is $\tau_p = 0.2$\,$\hbar/\Delta_\infty$.}
\end{figure}

In the previous section we have seen
how the oscillation of the ARPES intensity
can provide a deep insight into the creation process
of the collective Higgs oscillation of the order parameter.
Now, we will analyze the phase of these oscillations
at different points in momentum space
to resolve the deformation dynamic of the condensate symmetry.
We have seen that an $A_{1g}$-quench for $d$-wave
changes the sizes of the positive and negative lobes,
i.e. the lobes of one sign increase in size while the other decrease.
The resulting deformation of the condensate symmetry
will therefore be an oscillation,
where the smaller lobe increases while the larger lobe decreases.
Thus, the movement of the lobes is out of phase
(see Fig.~\ref{fig:phase_analysis}(a)).

The same quench applied to a nodal $s$-wave superconductor
with $f_\vk = \cos^2(2\varphi)$,
which has the same nodal structure,
will change its lobes all equally due to the same sign of the gap.
The resulting oscillations are therefore in phase
(see Fig.~\ref{fig:phase_analysis}(b)).
For comparison,
the quenched symmetry function and the resulting ARPES intensities
analogous to the $d$-wave case in Fig.~\ref{fig:dwave_quench}
can be found in the appendix in Fig.~\ref{fig:se_quench}.
Even though the oscillation of the lobes is in phase for this quench,
the gap oscillation shows a two peak structure (Fig.~\ref{fig:gap_osci}(d))
due to a shift of nodal lines.
Thus, a comparison of the Higgs oscillation spectrum
does not allow to distinguish between $d$-wave and nodal $s$-wave in this case.

To see whether we can observe the different condensate symmetry deformation
for these two gap symmetries in the ARPES intensity,
we extract the amplitude oscillations $\mathcal A(t,k=k_\mathrm F,\varphi)$
at $\varphi = 0$ and $\varphi = \pi/2$,
i.e. at points in momentum space rotated by $\varphi = \pi/2$,
which should oscillate with opposite phase in the case of $d$-wave
and with the same phase for the nodal $s$-wave.
The result can be found in Fig.~\ref{fig:phase_analysis}(c) and (d)
and confirms the expected oscillatory behavior.
Oscillations of the amplitude
provide therefore not only information about the amplitude of the gap
but relative phase information
from different points in momentum space can be extracted as well.
This shows that the information obtainable
from the time-dependent amplitude of the ARPES intensity
allows to clearly trace the induced symmetry deformation
and dynamic of the condensate
in momentum space,
e.g. to determine the movement of the $d$-wave or nodal $s$-wave lobes.

\section{Summary and Discussion}
In this article we demonstrate
how the collective Higgs oscillations of the superconducting order parameter
can be traced back and be observed
as oscillations of the ARPES intensity measured in THz tr-ARPES experiments.
There are two kinds of dynamics,
namely momentum-independent oscillations of the EDC maximum $\mathcal E(t)$,
which directly correspond to oscillations of the energy gap $\Delta(t)$,
and momentum-dependent oscillations of the amplitude
of the ARPES intensity $\mathcal A(t,k,\varphi)$.
While the EDC dynamic only reveals information
about the momentum-averaged quantity $\Delta(t)$,
the amplitude oscillation of the ARPES intensity
allows to perform a momentum-resolved analysis of the condensate dynamic.
Experimental conditions might dictate
which quantity can be observed more easily.
For this, further studies are required
including realistic matrix element effects for the ARPES intensity
as well as additional scattering channels
influencing the amplitudes of the oscillations.
What can be stated is
that both quantities require short enough probe pulses
to resolve the Higgs oscillations.
The EDC maximum oscillations are restricted in their amplitude
by the oscillation amplitude of the gap,
which can be in the sub-percentage range
depending on quench strength and gap symmetry.
Increasing the quench strength also increases the amplitude oscillations
up to a certain point beyond which the gap is quenched into a gapless regime.
Additional heating effects, which are not considered here,
restrict the maximum fluence of a quench pulse further,
such that only small quenches can be implemented experimentally.
In contrast, our calculation shows
that the amplitude oscillations of the ARPES intensity
can be up to several percentages in size also for small quench strength
and thus, be much larger for certain conditions.
However, they might be more difficult to resolve in general
as oscillations in energy.
Another difference can be found in the intrinsic damping.
While the oscillations in $\mathcal E(t)$ are intrinsically damped
due to dephasing,
the oscillations in $\mathcal A(t,k,\varphi)$ are undamped.
Yet, as the intrinsic damping follows only a power law,
additional exponential damping will diminish this advantage
when further scattering effects occur in an experiment.

A tracking of the amplitude oscillation frequencies $\omega_\vk^\mathcal A$
of the ARPES intensity $\mathcal A(t,k,\varphi)$
at different momenta leads to a deeper understanding
of the microscopic creation process
of collective Higgs oscillations $\Delta(t)$.
While the frequencies can be described
with a single formula independent of gap- and quench symmetry,
the relative weight of each frequency
contributing in the collective amplitude oscillation of the order parameter
is determined by the gap- and quench-symmetry in a nontrivial way.
If there are asymmetries in the weights,
additional Higgs modes can appear,
which are dynamically created due to quenching the condensate
in a symmetry channel different from its equilibrium value.

Furthermore, quenching the condensate in defined symmetry channels
and observing the oscillation phase of the ARPES intensity
allows to resolve the symmetry deformation and dynamic of the condensate.
This condensate dynamic depends on the gap and quench symmetry.
In this paper,
the quench symmetry for a certain gap was given
such that the resulting symmetry deformation dynamic of the condensate,
like an in-phase or out-of-phase oscillation of lobes of a $d$-wave
or nodal $s$-wave superconductor are clearly defined.
Importantly, this explicit construction of a quench symmetry
allows to distinguish between the two different gaps
with same nodal structure but different phase.

Whether and how all considered quenches can be realized experimentally
is an open question and beyond the scope of this paper;
however, there are several promising possibilities.
In-plane excitations can lead to symmetry breaking
due to the small but non-vanishing photon momentum \cite{NatCommun.11.287},
which allows to quench the system asymmetrically
with respect to its ground state symmetry.
Such momentum transfer can be enhanced and controlled in more detail
in transient grating setups
\cite{Science.300.1410,JChemPhys.120.4755}
or four-wave mixing experiments
\cite{PhysRevLett.109.147403}.
Once tailored quenches are realized experimentally,
a new spectroscopic tool will be available
to gain phase-sensitive information about the gap symmetry
of unconventional superconductors.

\begin{acknowledgments}
We thank J. Freericks for helpful comments
concerning the calculation of the ARPES intensity.
We are grateful to S. Kaiser for fruitful discussions.
We also thank N. Cheng and H. Krull for initial numerical studies.
We further thank the Max-Planck-UBC-UTokyo Center for Quantum Materials
for collaborations and financial support.
\end{acknowledgments}

\appendix

\section{Equations of motion}
\label{sec:eom}
Heisenberg's equation of motion Eq.~\eqref{eq:heom}
for the time-dependent quasiparticle operators
$\alpha_\vk^\dagger(t)$ and $\beta_\vk^\dagger(t)$
can be evaluated by using the ansatz in Eq.~\eqref{eq:ieom_ansatz}.
From the resulting equations,
a comparison of coefficients leads to a set of coupled
finite hierarchy differential equations
for the prefactors
\begin{widetext}
\begin{subequations}
\begin{align}
    &\partial_t a_{0\vk}(t) = \frac\im\hbar\Bigg[
        R_\vk(t)\Big[
            a_{0\vk}(t)\Big(|a_{0\vk}(t)|^2 + |a_{1\vk}(t)|^2
            - |b_{1\vk}(t)|^2\Big)
            - a_{1\vk}(t)b_{0\vk}(t)b_{1\vk}^*(t)
        \Big]
    \notag\\&\qquad\qquad\qquad
        + C_\vk(t)\Big[
            a_{0\vk}(t)\Big(a_{0\vk}(t)b_{1\vk}(t)
            + a_{1\vk}(t)b_{0\vk}(t)\Big)
        \Big]
        + C_\vk^*(t)\Big[
            b_{1\vk}^*(t)\Big(|a_{0\vk}(t)|^2 + |a_{1\vk}(t)|^2\Big)
        \Big]
    \Bigg]\,,
\\
    &\partial_t a_{1\vk}(t) = \frac\im\hbar\Bigg[
        - R_\vk \Big[
            a_{1\vk}(t)\Big(|b_{0\vk}(t)|^2 - |a_{1\vk}(t)|^2
            - |a_{0\vk}(t)|^2\Big)
            + a_{0\vk}(t)b_{0\vk}^*(t)b_{1\vk}(t)
        \Big]
    \notag\\&\qquad\qquad\qquad
    + C_\vk(t)\Big[
            a_{0\vk}(t)a_{1\vk}(t)b_{1\vk}(t) + a_{1\vk}(t)^2b_{0\vk}(t)
        \Big]
        + C_\vk^*(t)\Big[
            b_{0\vk}^*(t)\Big(|a_{0\vk}(t)|^2 + |a_{1\vk}(t)|^2\Big)
        \Big]
    \Bigg]\,,
\\
    &\partial_t b_{0\vk}(t) = \frac\im\hbar\bigg[
        R_\vk(t)\Big[
            b_{0\vk}(t)\Big(|b_{0\vk}(t)|^2 + |b_{1\vk}(t)|^2
            - |a_{1\vk}(t)|^2\Big)
            - a_{0\vk}(t)a_{1\vk}^*(t)b_{1\vk}(t)
        \Big]
    \notag\\&\qquad\qquad\qquad
        - C_\vk\Big[
            a_{1\vk}(t)b_{0\vk}(t)^2 + a_{0\vk}(t)b_{0\vk}(t)b_{1\vk}(t)
        \Big]
        - C_\vk^*\Big[
            a_{1\vk}^*(t)\Big(|b_{0\vk}(t)|^2 + |b_{1\vk}(t)|^2\Big)
        \Big]
    \bigg]\,,
\\
    &\partial_t b_{1\vk}(t) = \frac\im\hbar\bigg[
        - R_\vk(t)\Big[
            b_{1\vk}(t)\Big(|a_{0\vk}(t)|^2 - |b_{0\vk}(t)|^2
            - |b_{1\vk}(t)|^2\Big)
            + a_{0\vk}^*(t)a_{1\vk}(t)b_{0\vk}(t)
        \Big]
    \notag\\&\qquad\qquad\qquad
        - C_\vk\Big[
            a_{1\vk}(t)b_{0\vk}(t)b_{1\vk}(t) + a_{0\vk}(t)b_{1\vk}(t)^2
        \Big]
        - C_\vk^*(t)\Big[
            a_{0\vk}^*(t)\Big(|b_{0\vk}(t)|^2 + |b_{1\vk}(t)|^2\Big)
        \Big]
    \bigg]\,.
\end{align}
\label{eq:eom_prefactors}%
\end{subequations}
\end{widetext}
The initial values for the prefactors can be derived
from the initial values of the quasiparticle expectation values.
The time-dependent quasiparticle expectation values
expressed within the iterated equation of motion ansatz
assuming that the system is in equilibrium for $T=0$ read
\begin{subequations}
\begin{align}
    \braket{\alpha_\vk^\dagger(t)\alpha_{\vk}(t')}
    &= a_{1\vk}(t)a_{1\vk}^*(t')\,,\\
    \braket{\beta_\vk^\dagger(t)\beta_{\vk}(t')}
    &= b_{1\vk}(t)b_{1\vk}^*(t')\,,\\
    \braket{\alpha_\vk(t)\beta_{\vk}(t')}
    &= a_{0\vk}^*(t)b_{1\vk}^*(t')\,,\\
    \braket{\alpha_\vk^\dagger(t)\beta_{\vk}^\dagger(t')}
    &= a_{1\vk}(t)b_{0\vk}(t')\,.
\end{align}
\label{eq:td_exp_values_ieom}%
\end{subequations}
\begin{table}[b]
    \centering
    \setlength{\tabcolsep}{10pt}
    \begin{tabular*}{\linewidth}{ll}
    \toprule
    Parameter & Value\\
    \toprule
    Dispersion $\epsilon_\vk$ & $t \vk^2 - \epsilon_\mathrm F$\\
    Dispersion parameter $t$ & $2005.25 \,\mathrm{meV}$\\
    Fermi energy $\epsilon_\mathrm F$ & $9470 \,\mathrm{meV}$\\
    Gap value $\Delta$ & $1.35\,\mathrm{meV}$\\
    Energy cutoff $\epsilon_\mathrm c$ & $8.3\,\mathrm{meV}$\\
    Interaction quench strength $g_{\mathrm i}$ & $0.8$\\
    State quench strength $g_{\mathrm s}$ & $0.4$\\
    Temperature $T$ & $0\,\mathrm{K}$\\
    Radial discretization points $N_k$ & $1000$\\
    Azimuthal discretization points $N_\varphi$ & $1000$\\
    \bottomrule
    \end{tabular*}
    \caption{Parameters used in the numerical calculation.
    The values correspond to material parameters of lead, following
    \cite{PhysRevB.76.224522}.
    All results depend only quantitatively on the parameters
    and can be rescaled to any other material values.}
    \label{tab:params}
\end{table}%
To derive this result, we used the fact that for $T=0$,
all equilibrium quasiparticle expectation values,
i.e. $\braket{\alpha_\vk^\dagger\alpha_{\vk}} = 0$ etc. are zero.
For $t=t'=0$, we easily find
\begin{subequations}
\begin{align}
    a_{0\vk}(0) &= b_{0\vk}(0) = 1\,,\\
    a_{1\vk}(0) &= b_{1\vk}(0) = 0\,,
\end{align}
\label{eq:ai0bi0_eq}%
\end{subequations}
i.e. $\alpha_{\vk}^\dagger(0) = \alpha_\vk^\dagger$ and
$\beta_{\vk}^\dagger(0) = \beta_\vk^\dagger$.
The solution for a nonequilibrium distribution is straight forward as well,
by using the respective initial values for quasiparticle expectation values.
However, we found it advantageous for more numerical stability
to first use the relation
\begin{subequations}
\begin{align}
    |a_{0\vk}(t)|^2 + |a_{1\vk}(t)|^2 &= 1\,,\\
    |b_{0\vk}(t)|^2 + |b_{1\vk}(t)|^2 &= 1\,,
\end{align}
\end{subequations}
following from the anticommutation rules
for the time-dependent Heisenberg operators,
to rewrite Eq.~\eqref{eq:td_exp_values_ieom} as
\begin{subequations}
\begin{align}
    \braket{\alpha_\vk^\dagger\alpha_{\vk}}(0)
    &= 1 - |a_{0\vk}(0)|^2\,,\\
    \braket{\beta_\vk^\dagger\beta_{\vk}}(0)
    &= 1- |b_{0\vk}(0)|^2\,,\\
    \braket{\alpha_\vk\beta_{\vk}}(0)
    &= a_{0\vk}^*(0)b_{1\vk}^*(0)\,,\\
    \braket{\alpha_\vk^\dagger\beta_{\vk}^\dagger}(0)
    &= a_{1\vk}(0)b_{0\vk}(0)\,,
\end{align}
\end{subequations}
such that the solution reads
\begin{subequations}
\begin{align}
    a_{0\vk}(0) &= \sqrt{1-\braket{\aDa}(0)}\,,\\
    b_{0\vk}(0) &= \sqrt{1-\braket{\bDb}(0)}\,,\\
    a_{1\vk}(0) &= \frac{\braket{\aDbD}(0)}{\sqrt{1-\braket{\bDb}(0)}}\,,\\
    b_{1\vk}(0) &= \frac{\braket{\ab}^*(0)}{\sqrt{1-\braket{\aDa}(0)}}\,.
\end{align}
\label{eq:ai0bi0}
\end{subequations}
This prevents dividing by zero
in the calculation of $a_{1\vk}(0)$ and $b_{1\vk}(0)$.

The initial values $\braket{\aDa}(0)$ etc. in the state quenched case
are derived as follow.
For $T=0$, the equilibrium electron expectation values read
\begin{subequations}
\begin{align}
    \braket{c_{-\vk\downarrow}c_{\vk\uparrow}} &= \frac{\Delta_\vk}{2E_\vk}\,,\\
    \braket{c_{\vk\uparrow}^\dagger c_{-\vk\downarrow}^\dagger}
        &= \frac{\Delta_\vk^*}{2E_\vk}\,,\\
    \braket{c_{\vk\uparrow}^\dagger c_{\vk\uparrow}}
        &= \frac 1 2 - \frac{\epsilon_\vk}{2E_\vk}\,,\\
    \braket{c_{-\vk\downarrow} c_{-\vk\downarrow}^\dagger}
        &= \frac 1 2 + \frac{\epsilon_\vk}{2E_\vk}\,.
\end{align}
\end{subequations}
Using the definition of the Bogoliubov transformation
in Eq.~\eqref{eq:bt_transform}
one can derive the corresponding expressions
for the quasiparticle expectation values.
In the state quench,
all occurring symmetry functions $f_\vk$
are replaced by the modified symmetry function $f_\vk'$.
These expectation values are then used as initial values for the calculation.

\section{Influence of probe pulse width}
\label{sec:pulse_width}
\begin{figure}[t]
    \centering
    \includegraphics[width=\columnwidth]{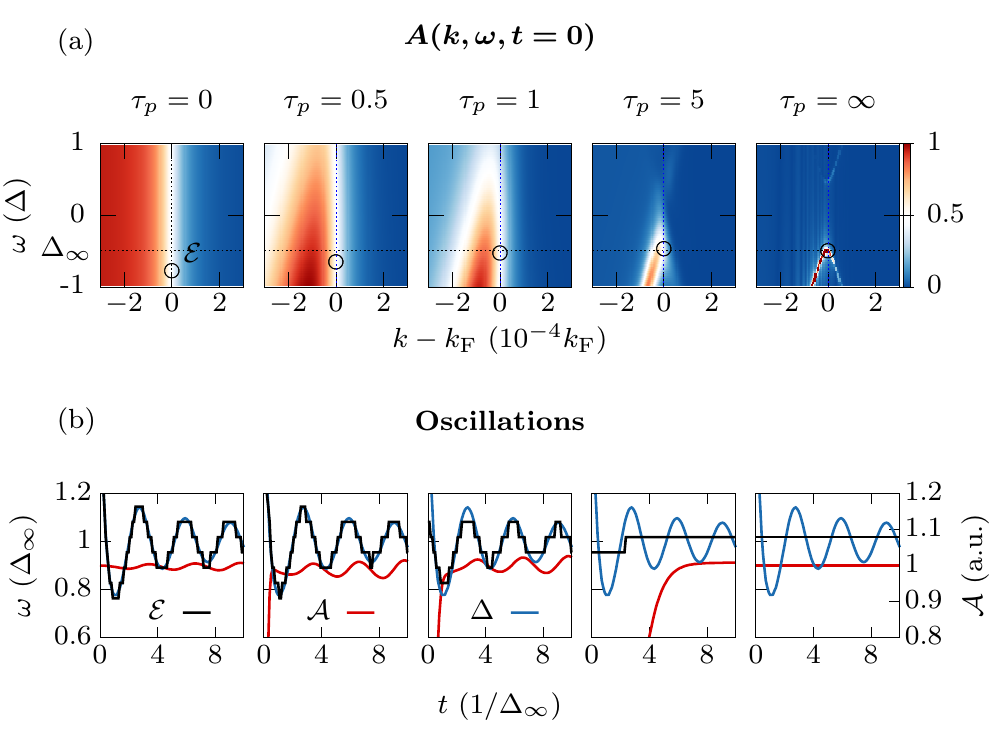}
    \caption{\label{fig:probe_pulse}%
    Influence of probe pulse width on ARPES spectrum
    and its time evolution for an s-wave superconductor
    after an interaction quench.
    For decreasing probe pulse width (right to left),
    the energy resolution decreases while the time-resolution increases.
    The edges in $\mathcal E$
    are an artifact due to the finite frequency resolution.
    The decreased amplitude of the ARPES intensity in the beginning
    for larger pulse widths $\tau_\mathrm p$
    results from decreased weight
    due to the cutoff of the probe pulse envelope at $t=0$.}
\end{figure}%
The probe pulse width determines both the energy resolution in the spectrum
and the time resolution in the dynamic \cite{PhysRevX.3.041033}.
If the probe pulse is very long
compared to the intrinsic oscillations of the system,
i.e. $\tau_p \gg \hbar/(2\Delta_\infty)$,
it can only detect the average value of the oscillations,
which results in constant quantities $\mathcal E$ and $\mathcal A$.
However, if the probe pulse is short in time,
i.e. shorter than the intrinsic oscillation period of the system,
it can scan the oscillations
which leads to oscillations in the ARPES intensity in energy and amplitude.

In Fig.~\ref{fig:probe_pulse},
the ARPES intensity of an $s$-wave superconductor
after an interaction quench is shown for different probe pulse widths.
An infinite long probe pulse with $\tau_p = \infty$
leads to a very sharp ARPES intensity in energy.
However, no oscillations in the ARPES intensity
can be observed in this case (black and red curve),
despite the oscillations in the gap (blue curve).
For decreasing probe pulse width,
the energy resolution of the ARPES intensity decreases
and the ARPES intensity peak becomes broader.
Simultaneously, the time-resolution increases
and oscillations of the EDC maximum $\mathcal E(t)$
and the ARPES intensity amplitude $\mathcal A(t)$ can be observed.

\section{State quench for nodal $s$-wave superconductor}
\label{sec:nod_s}
\begin{figure}[t]
    \centering
    \includegraphics[width=\columnwidth]{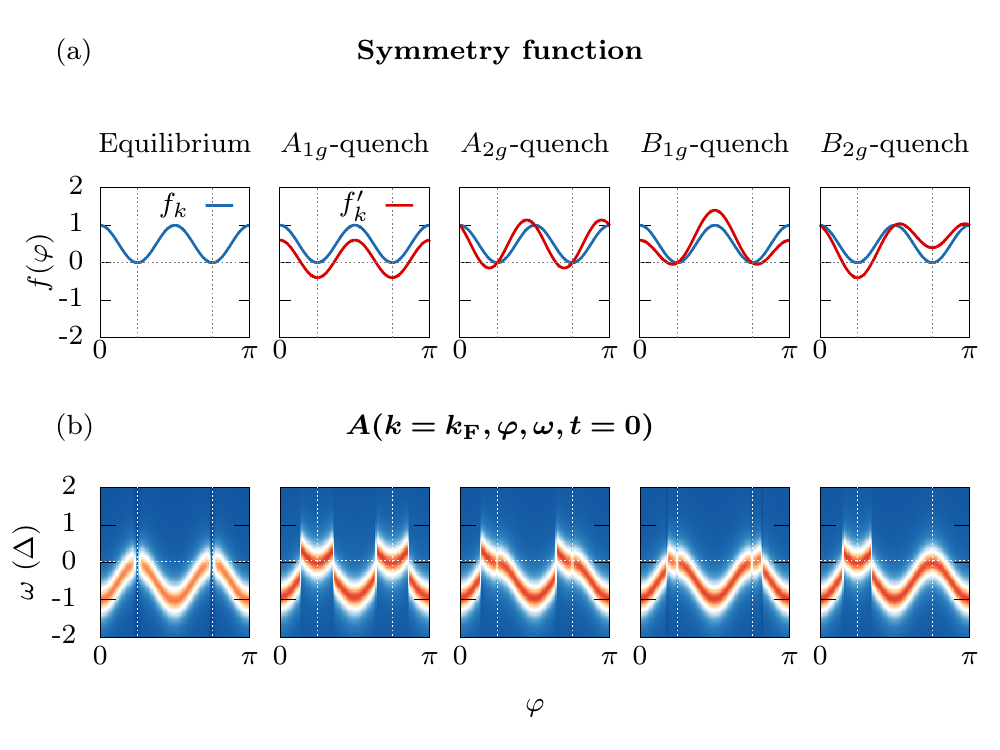}
    \caption{\label{fig:se_quench}%
    ARPES intensity of a nodal $s$-wave superconductor
    after a state quench.
    Top: Symmetry function $f_\vk$ (blue)
    and quenched symmetry function $f_\vk'$(red) for different channels.
    Bottom: Corresponding ARPES intensity,
    shift of nodal lines is reflected by a shift of spectral weight
    in the nodal region.
    The probe pulse width is $\tau_p = 10$\,$\hbar/\Delta_\infty$.}
\end{figure}
In analogy to Sec.~\ref{sec:quench_dwave} and Fig.~\ref{fig:dwave_quench},
we quench a nodal $s$-wave superconductor
in all symmetry channels of the $D_{4h}$ lattice point group.
The result can be found in Fig.~\ref{fig:se_quench}.
In the areas, where the quench shifts nodal lines,
we can observe a corresponding shift in spectral weight.
If one calculates the resulting gap oscillations
shown in Fig.~\ref{fig:gap_osci},
one can observe a two peak or kink structure for all quenches.
We can understand this by the fact that
all quenches shift the nodal lines
such that there is an asymmetry
in the weights of the $k$-dependent frequencies.
In the case of the $B_{1g}$-quench,
this shift is smaller compared to the other quenches,
which results in a lower frequency of the second Higgs mode.
In general, the energy of the dynamically created Higgs mode
depends on the quench strength,
i.e. the deviation from the equilibrium state.

\section{Momentum-dependent frequency}
\label{sec:mdep_freq}
If we are only interested in the dynamic of the gap
and expectation values at equal times,
we can formulate the BCS Hamiltonian in Anderson's pseudopin representation
\cite{PhysRev.112.1900}
and solve the resulting Bloch equations similar to
\cite{NatCommun.11.287}.
The pseudospin is defined as
\begin{align}
    \vec{\sigma}_\vk &=
        \frac{1}{2} \Psi_\vk^\dagger \vec{\tau} \Psi_\vk
\end{align}
with the Nambu-Gorkov spinor
$\Psi_{\vec{k}}^\dagger = %
\begin{pmatrix} c_{\vk\uparrow} & c^\dagger_{-\vk\downarrow} \end{pmatrix}$
and the Pauli matrices $\tau^i$.
The $x$-component corresponds to the real part
of the anomalous Green's function
$\braket{\cc}+\braket{\cDcD}$,
while the $z$-component corresponds to the normal Green's function
$\braket{\cDc}-\braket{\ccD}$.
The BCS Hamiltonian can be written as
\begin{align}
    H(t) &= \sum_\vk \vec{b}_\vk(t) \vec{\sigma}_\vk
\end{align}
with the pseudomagnetic field
$\vec{b}_\vk(t) = \left( -2\Delta(t) f_\vk , 0 , 2\epsilon_\vk\right)$
and the gap equation reads
\begin{align}
    \Delta(t) &= V \sum_\vk f_\vk \braket{\sigma_\vk^x}(t)\,.
\end{align}
Hereby, we assume a real gap, such that the $y$-component,
which corresponds to the imaginary part of the gap, is zero.
Heisenberg's equations of motion for the pseudospin
take the form of Bloch equations
\begin{align}
    \partial_t \braket{\vec{\sigma}_\vk}(t)
        &= \vec{b}_\vk(t) \times \braket{\vec{\sigma}_\vk}(t)\,.
\end{align}
We can linearize these equations for small deviations from the initial values
\begin{subequations}
\begin{align}
    \braket{\vec \sigma_\vk}(t)
        &= \braket{\vec \sigma_\vk}(0)
            + \braket{\delta \vec \sigma_\vk}(t)\,, \\
    \Delta(t) &= \Delta(0) + \delta \Delta (t)\,,
\end{align}
\end{subequations}
where
$\braket{\vec \sigma_\vk}(0)$
is the quenched pseudospin expectation value at $t=0$ with
\begin{subequations}
\begin{align}
    \braket{\sigma_\vk^x}(0) &= \braket{\sigma_\vk^x}'
        = \frac{\Delta_\vk'}{2E_\vk'}\,,\\
    \braket{\sigma_\vk^z}(0) &= \braket{\sigma_\vk^z}'
        = -\frac{\epsilon_\vk}{2E_\vk'}
\end{align}
\end{subequations}
according to Sec.~\ref{sec:quenches} in the main text.
Using this ansatz while neglecting
all products in higher order of the deviations,
we can write the Bloch equations in Laplace space for complex frequency $s$ as
\begin{subequations}
\begin{align}
    s \braket{\delta \sigma_\vk^x} (s)
        &= -2 \epsilon_\vk \braket{\delta \sigma_\vk^y}(s)\,, \\
    s \braket{\delta \sigma_\vk^y} (s)
        &= \frac{2 \epsilon_\vk}{s}
            \frac{\Delta f'_\vk-\Delta(0)f_\vk}{2E'_\vk} \nonumber \\
        &\quad+ 2 \epsilon_\vk\braket{\delta \sigma_\vk^x}(s)  \nonumber \\
        &\quad+ 2 \Delta(0) f_\vk \braket{\delta \sigma_\vk^z}(s) \nonumber \\
        &\quad- 2 \epsilon_\vk \frac{f_\vk}{2E'_\vk}\delta \Delta (s)\,, \\
    s \braket{\delta \sigma_\vk^z} (s)
        &= - 2 f_\vk \Delta(0) \braket{\delta \sigma_\vk^y}(s)\,.
\end{align}
\label{eq:bloch_linear}
\end{subequations}
Solving for the $x$- and $z$-component, it follows
\begin{subequations}
\begin{align}
    \braket{\delta \sigma_\vk^x}(s) &=
        \frac{2\epsilon^2 E_\vk'
            ( \frac{\Delta(0)f_\vk - \Delta f_\vk'}{s} + f_\vk \delta\Delta(s))
        }
        {4\epsilon^2 + 4\Delta(0)^2f_\vk^2 + s^2}\,,\\
    \braket{\delta \sigma_\vk^z}(s) &=
        \frac{2\epsilon \Delta(0) f_\vk E_\vk'
            ( \frac{\Delta(0)f_\vk - \Delta f_\vk'}{s} + f_\vk \delta\Delta(s))
        }
        {4\epsilon^2 + 4\Delta(0)^2f_\vk^2 + s^2}
\end{align}
\label{eq:dxdz}
\end{subequations}
and the solution for the gap reads
\begin{align}
    \delta\Delta(s) &= \frac{F_2(s)}{1-F_1(s)}
\end{align}
with
\begin{widetext}
\begin{subequations}
\begin{align}
    F_1(s) &= C
        - V \sum_\vk f_\vk^2
        \frac{s^2 + 4 \Delta(0)^2 f_\vk^2}
        {2E'_\vk \left(
            s^2 + 4 \epsilon_\vk^2 + 4 \Delta(0)^2 f_\vk^2
        \right)}\,, \\
    F_2(s) &= \frac{\Delta(0)}{s}\left(C-1\right)
        + \frac{1}{s}
        V \sum_\vk
            f_\vk (\Delta f'_\vk - \Delta(0) f_\vk )
            \frac{s^2 + 4\Delta(0)^2 f_\vk^2}
                {2E'_\vk(s^2+4\epsilon_\vk^2 + 4\Delta(0)^2 f_\vk^2)}\,.
\end{align}
\end{subequations}
with $C = V\sum_\vk \frac{f_\vk^2}{2E'_\vk}$.
An inverse Laplace transform yields
\begin{subequations}
\begin{align}
    F_1(t) &= C
        - V \sum_\vk f_\vk^2 \Big(
            \frac{\delta(t)}{2E_\vk'}
            - \frac{2\epsilon_\vk^2}{E_\vk' \omega_\vk} \sin(\omega_\vk t)
        \Big)\,,\\
    F_2(t) &= \Delta(0)\left(C-1\right)
        + V \sum_\vk
            f_\vk (\Delta f'_\vk - \Delta(0) f_\vk )
            \Big( \frac{2\Delta(0)^2 f_\vk^2
                + \epsilon_\vk^2 \cos(\omega_\vk t)}{E_\vk' \omega_\vk^2}
            \Big)\,,
\end{align}
\end{subequations}
where $\omega_\vk = 2\sqrt{\epsilon_\vk^2 + \Delta(0)^2 f_\vk^2}$.
\end{widetext}
In the linear regime, we can approximate $\Delta(0) \approx \Delta_\infty$,
such that we arrive at the formula Eq.~\eqref{eq:winf} from the main text.
We can see that the main oscillation frequency
of the summand in the gap equation
and both the anomalous and normal Green's function,
i.e. the $x$- and $z$-component of the pseudospin in Eq.~\eqref{eq:dxdz}
is given by $\omega_\vk^\mathcal A$.
The frequency is independent of the quench,
however the individual weighting in momentum space,
i.e. the amplitudes of each frequencies depend heavily on the quench.
Namely, the prefactors, controlled by the gap symmetry $f_\vk$
and quenched symmetry $f_\vk'$,
determine the dominating frequency
in the summation process in a nontrivial way.

We can see the energy scales of the collective Higgs modes in more detail,
by evaluating the $k$-sum in $F_1(s)$ or $F_2(s)$.
For this,
the sum is replaced by an integral using polar coordinates, i.e.
$V\sum_\vk \rightarrow \lambda \int_{-\infty}^\infty \,\mathrm d\epsilon
\,\int_0^{2\pi}\,\mathrm d\varphi$,
where $\lambda = VD(\epsilon_\mathrm F)$
with $D(\epsilon_\mathrm F)$ the density of states at the Fermi level.
Following \cite{NatCommun.11.287},
one finds
\begin{align}
    F_1(s) \propto \int_0^{2\pi} \frac{\sqrt{s^2 + 4\Delta(0)^2f(\varphi)^2}}{\sqrt{s^2 + 4(\Delta(0)^2f(\varphi)^2 - \Delta^2f'(\varphi)^2)}}
\end{align}
and a similar expression for $F_2(s)$.
Here,
two energy scales appear.
The term $\sqrt{s^2 + 4\Delta(0)^2f(\varphi)^2}$
determines the usual $2\Delta$ Higgs mode,
whereas the term
$\sqrt{s^2 + 4(\Delta(0)^2f(\varphi)^2 - \Delta^2f'(\varphi)^2)}$
determines the energy of the second low-lying Higgs mode.
For small quenches $g\approx 0$,
the quenched symmetry function is $f'(\varphi) \approx f(\varphi)$
and the second Higgs mode softens.
For increasing quench strength,
the difference $\Delta(0)^2f(\varphi)^2 - \Delta^2f'(\varphi)^2$
increases, resulting in an increase of the energy.
In contrast,
the $2\Delta$ Higgs mode decreases as the gap is reduced more.
Yet, to obtain the exact energies,
the integral has to be evaluated numerically
as the whole expression is nontrivial.

The gap oscillations are damped with $1/t^n$,
where $n=1/2$ for $s$-wave and $n> 1/2$ for $d$-wave.
This is a dephasing effect
as oscillations of slightly different frequency are summed up
in the gap equation.
In contrast, the oscillations of the quasiparticles or pseudospins,
which are the quantities which are summed up in the gap equation,
are undamped.
This can be seen in the expression for the pseudospins in Eq.~\eqref{eq:dxdz}.
As an example the expression for the $z$-component reads
\begin{align}
    \braket{\sigma_\vk^z}(s) &= \frac{a}{s(b^2+s^2)}
        + \frac{c}{b^2+s^2}\delta\Delta(s)\,,
\end{align}
where $a$, $b$, $c$ are $s$-independent constants.
The inverse Laplace transform of the first term yields
$a/b^2 - a \cos(b t)/b^2$, which contains an undamped oscillatory part.
As a result,
the considered amplitude oscillations $\mathcal A(t,k,\varphi)$
in this work are undamped as well.

\section{Influence of system parameters}
\label{sec:influence_parameters}
Higgs oscillations are a collective oscillation of the order parameter.
Thus, a variation of system parameters
only quantitatively changes the oscillations
but do not affect the overall collective excitation.
To demonstrate that the induced Higgs oscillation
do not depend on system parameters,
we show results for a single-band tight-binding dispersion on a square lattice,
where we vary the filling and quench strength.
The dispersion reads
\begin{align}
    \epsilon_\vk &= -2t(\cos k_x + \cos k_y) - \epsilon_\mathrm F
    \label{eq:tb_dispersion}
\end{align}
where we chose $t = 200$\,meV, $\Delta = 30$\,meV,
$\epsilon_\mathrm c = 60$\,meV, $N_k = 2000$ and $N_\varphi = 1000$,
which is in the order of numerical values found e.g. in cuprates.

\begin{figure}[t]
    \centering
    \includegraphics[width=\columnwidth]{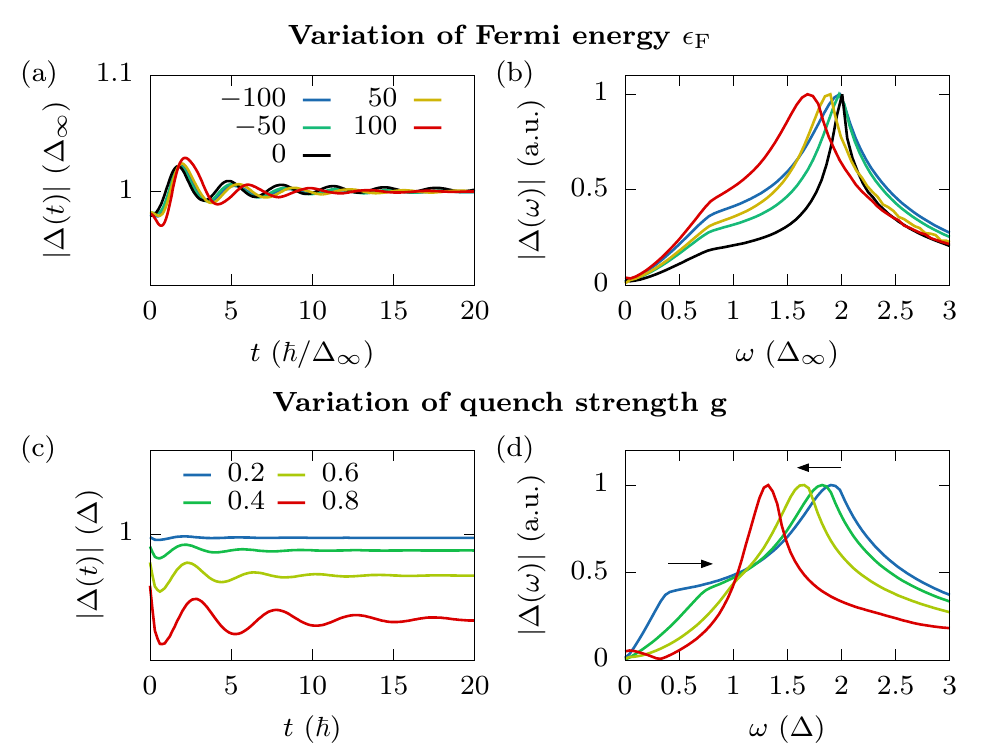}
    \caption{\label{fig:gap_osci_var_dispersion}%
    Higgs oscillations and spectrum for a $d$-wave order parameter
    quenched in the $A_{1g}$ channel for varying parameters
    using the tight-binding dispersion Eq.~\eqref{eq:tb_dispersion}.
    (a) and (b) The Fermi energy $\epsilon_\mathrm F$ is varied
    from negative to positive values,
    which results in a change in the Fermi surface form nearly circular
    to discontiguous Fermi pockets.
    In all cases, 2 Higgs modes occur,
    yet, the relative weight of the occurring frequencies
    depends on the Fermi surface.
    (c) and (d) The quench strength $g$ is varied.
    For increasing quench strength,
    the $2\Delta_\infty$ Higgs mode is shifted to lower energies,
    while the low-lying Higgs mode is shifted to higher energies.
    }
\end{figure}

In Fig.~\ref{fig:gap_osci_var_dispersion}(a) and (b),
we show the results for a $d$-wave order parameter
which we quench in the $A_{1g}$ channel with a fixed strength of $g=0.4$.
We vary the filling by varying the Fermi energy
from $\epsilon_\mathrm F = -100$\,meV to $\epsilon_\mathrm F = 100$\,meV.
This changes the Fermi surface from nearly circular
to discontiguous Fermi pockets at the diagonals.
First of all,
independent of the parameters,
Higgs oscillations are always excited as shown in
Fig.~\ref{fig:gap_osci_var_dispersion}(a)
and in all cases, two Higgs modes are visible as it can be seen in
Fig.~\ref{fig:gap_osci_var_dispersion}(b).
As the Fermi surface changes for different fillings,
the same quench strength $g$ has a different strong quench effect
on the system, yet,
the result is qualitatively the same.
As we have shown in this paper
the ARPES intensity directly reflects the collective gap oscillations
and the oscillations of the underlying condensate.
Thus,
the qualitative invariance of the gap oscillations
under variation of system parameters is also found in the oscillations
of the ARPES intensity.

Quantitatively, the following differences are visible.
In the half-filling case $\epsilon_\mathrm F = 0$,
where the Fermi surface has the shape of a square
with the corners in the antinodal direction,
the $2\Delta_\infty$ peak is sharper compared to the other cases.
This is perfectly understandable with our analysis of the main text
regarding the emerging of the Higgs modes.
Due to the position of the corners in the antinodal direction,
there is more weight in this region
and the corresponding frequencies contribute more.
As the frequency in the antinodal direction at the Fermi level
is given by the full open gap $2\Delta_\infty$,
this frequency dominates the resulting spectrum.

Another difference is visible
for discontiguous Fermi surfaces with $\epsilon_\mathrm F > 0$.
Here,
there is no Fermi surface in the antinodal direction and
thus, less weight for the $2\Delta_\infty$ oscillation.
As a consequence,
the main peak is slightly shifted to lower frequencies.

In Fig.~\ref{fig:gap_osci_var_dispersion}(c) and (d),
we fix the filling at $\epsilon_\mathrm F = -100$\,meV
and vary the quench strength using $g=[0.2,0.4,0.6,0.8]$.
We observe
that the gap is reduced more for stronger quenches,
i.e. $\Delta_\infty$ decreases.
As a consequence,
the main oscillation frequency of the $2\Delta_\infty$ Higgs mode decreases as well,
which can already be seen in the oscillations
in Fig.~\ref{fig:gap_osci_var_dispersion}(c)
but more clearly in Fig.~\ref{fig:gap_osci_var_dispersion}(d)
by a shift of the main peak to lower frequencies.
In contrast, as also discussed in appendix~\ref{sec:mdep_freq},
the energy of the low-lying Higgs mode increases.

\end{document}